\documentclass{article}
\usepackage[utf8]{inputenc}

\title{Pragmatic Information in Quantum Mechanics}
\author{Juan G. Roederer \footnote{Permanent address: Frasier Meadows, 4825 Sioux Drive, Apt 102, Boulder, Colorado 80303, USA. URL: www.gi.alaska.edu/$\sim$Roederer}\\Geophysical Institute, University of Alaska Fairbanks\\Fairbanks AK, 99775, USA}

\begin{document}

\maketitle

\textit{Abstract}: An objective definition of pragmatic information and the consideration of recent results about information-processing in the human brain can help overcome some traditional difficulties with the interpretation of quantum mechanics. Rather than attempting to define information \textit{ab initio}, I introduce the concept of interaction between material bodies as a primary concept. Two distinct categories can be identified: 1) Interactions which can always be reduced to a superposition of physical interactions (forces) between elementary constituents; 2) Interactions between complex bodies which cannot be reduced to a superposition of interactions between parts, and in which \textit{patterns and forms} (in space and/or time) play the determining role. Pragmatic information is then defined as the correspondence between a given pattern and the ensuing pattern-specific change. I will show that pragmatic information is a biological concept that plays no active role in the purely physical domain---it only does so when a living organism intervenes. The consequences for the foundations of both classical and quantum physics are important and will be discussed in detail. Since by its very definition pragmatic information---the one our brain uses to represent, think about and react to the outside world---cannot operate in the quantum domain, it is advisable to refrain from using it in our mental representation of what is happening ``inside'' a quantum system. Although the theoretical framework developed for quantum mechanics handles mathematical entities specifically attributed to a quantum system, the only truly \textit{pragmatic} information it can provide relates to macroscopic effects on the environment (natural, or artificial as in a measurement instrument) with which the system interacts.  

\section{What is Pragmatic Information?}

We live in the Information Era. It is quite understandable that a term of such every-day usage as ``information'' remains largely undefined when used in the general scientific literature, despite the fact that philosophers, mathematicians, linguists, engineers, biologists and writers use this common term with quite different and distinct meanings. Physicists are accustomed to working with the mathematically well-founded concepts of Shannon (or statistical) and algorithmic information. These terms are used to designate objective, quantitative expressions of the \textit{amount} of information in a message; the \textit{gain} of information when alternatives are resolved; the \textit{degree} of uncertainty; the \textit{quality} of information; the minimum \textit{number} of binary steps to identify or describe something; the maximum number of bits that may be processed in one location; the total number of bits available in the Universe; etc. However, whenever physicists use the term information in statements in a qualitative, descriptive way such as ``the field carries information about its sources''; ``information cannot travel faster than light''; ``information about the microstate of a gas''; ``the detector provides information about the radiation background''; ``information deposited in (or extracted from) the environment''; etc., they are really talking about \textit{pragmatic information}. 

In effect, it is mostly the biologists, particularly geneticists and neuroscientists, who make an explicit distinction between Shannon or statistical information (detached from purpose or effect), algorithmic information (minimum number of binary steps to define something, or the numerical values of physical variables) and pragmatic information (linking purpose and effect) (e.g., see Küppers 1990). Of these three classes of information the last one is usually the most relevant in biology; indeed, the notion of \textit{quantity} of information is often of secondary importance: what counts is what information ultimately \textit{does}, not how many bits are involved.

As we argue in Roederer (2005), pragmatic information is a purely biological concept. This seems quite difficult to accept for physicists, not the least because many agree with John Wheeler's dictum that ``every physical quantity derives its ultimate significance from bits, binary yes-or-no indications'' (Wheeler 1989). Yet let me emphasize the trivial facts that physics is the product of information-processing by \textit{brains}, based on interactions of systems chosen and prepared by \textit{brains}, measured with artifacts (instruments) created by \textit{brains}, using algorithms (mathematics and models) developed by \textit{brains} and following plans, purposes and quantitative predictions made by \textit{brains}. Perhaps the principal obstacle in an effort to persuade physicists to accept the idea of information being a purely biological concept, is the perceived lack of a widely accepted definition of pragmatic information that is truly objective, i.e., unrelated to any human use or practice.\footnote{From its beginning, traditional information theory (Shannon and Weaver 1949) deliberately refrained from giving a universal and objective definition of the concept of information \textit{per se}.}

Philosophers have struggled for a long time with an \textit{ab initio} definition of information. Physicists often consider information in its general sense to be a statement that resolves a set of alternatives, i.e., which resolves uncertainty. Others consider ``data'' to be the essence of information. Most visualize information as a quantity expressed in number of bits. Rather than attempting to define information \textit{ab initio}, I find it more appropriate to start with \textit{interaction between bodies} as the primary concept or ``epistemological primitive'' (Roederer 1978, 2005), and from there derive the concept of information. We can identify two distinct groups. Category 1) interactions can always be reduced to a linear \textit{superposition} of mutual physical interactions (i.e., forces) between the interacting bodies' elementary constituents, in which energy transfer between the interacting parts plays a fundamental role. Category 2) interactions cannot be expressed as a superposition of elementary interactions, and it is \textit{patterns and forms} (in space and/or time) that play the determining role on whether or not an interaction is to take place; the required energy must be provided from outside through a specific complex \textit{interaction mechanism}. 

Examples of category 1), which we call \textit{force-driven} interactions, are all the physical interactions between elementary particles, wave fields, nuclei, atoms, molecules, parcels of fluid, complex solid bodies and networks, planets and stars. Ultimately, they all originate in the four basic interactions between fundamental particles (electromagnetic, gravitational, strong and weak). It is the fundamental property of  reducibility of force-driven interactions, that allows physics to work with approximate \textit{models} of the complex reality ``outside'' to make quantitative predictions or retrodictions about the time-evolution of a given physical system. 

The simplest case of an interaction of category 2) is any arrangement in which the presence of a specific pattern in a complex system $S$ (sometimes called the sender or the source) leads to a causal, macroscopic and \textit{univocal} change in another complex system $R$ (the recipient), a change that would not happen (or just occur by chance) in the absence of the particular pattern at the source. Typical examples range from effects on their respective chemical environments of the one-dimensional pattern of bases in the RNA molecule or the three-dimensional shape of a folded protein; the light patterns detected by an insect and the resulting shape of its flight orbit around a light source; the patterns of neural electrical impulses in one region of the brain triggering impulses in another; to the perceived print patterns changing the state of knowledge in a reader's brain. It is important to point out that information-driven interactions all require a complex interaction mechanism with a reset function (often considered part of one of the interacting bodies), and which ultimately provides the energy required to effect the specific change. Although we call them interactions, they are causally unidirectional, from source to recipient. However, the designation \textit{inter}-action is justified in the sense that, to occur, they require some predesigned or evolutionary match (sometimes called ``understanding'') between source and recipient.

Pragmatic information is then defined as \textit{that which represents the univocal correspondence pattern $\rightarrow$ change}; it is the reason why we call Category 2) \textit{information-driven interactions} (Roederer 2003).\footnote{This may appear to some readers as a circular definition. It is not: ``information-driven'' is just a convenient name for this category; we could have called them ``pattern-driven interactions''.} By ``univocal'' we mean that the interaction process is deterministic and must yield identical results when repeated under similar conditions of preparation\footnote{We shall ignore at this stage considerations of matching errors, fluctuations. etc.}. This in turn means that the triggering pattern must be stable during some finite time and be amenable to be copied. When an information-driven interaction has occurred, we say that pragmatic information was transferred from the source $S$ (where the pattern resides) to the recipient $R$ (where the pattern-specific change occurs). We wish to emphasize that in this definition, the concept of pattern refers to a physical/topological property in space and/or instants of time;\footnote{In philosophy there is a large literature on \textit{semiotics}, the study of signs, their meaning and effects, first developed by C. S. Peirce in the late 19th century.} it includes, but is not limited to, symbols to which one can assign syntactic and semantic dimensions (e.g., see K\"uppers, 1990, Chapter 3). 

Note that in the preceding we do not imply that information ``resides'' in the patterns---the concept of pragmatic information is one of \textit{relationship} between patterns and changes, mediated by some interaction mechanism. A pattern all by itself has no meaning or function. On occasion one may say that information is \textit{encoded} in the pattern at the source. The effected change can itself be a pattern. Finally, there is no such thing as a numerical \textit{measure} of pragmatic information. Pragmatic information cannot be quantified---it represents \textit{a correspondence} which either exists or not, or works as intended or not, but it cannot be assigned a magnitude. There are cases, of course, where a given pattern and/or the change can be expressed in numbers or bits by an external agent, but that number \textit{is not} the pragmatic information involved. 

There are only three fundamental processes through which mechanisms of information-driven interactions can emerge (Roederer 2005), involving processes at three vastly different time scales: 1) Darwinian evolution; 2) adaptation or neural learning; 3) as the result of human reasoning and long-term planning. In other words, they all involve \textit{living matter}---indeed, information-driven interactions represent the \textit{defining property of life} (Roederer, 2004). In this overall evolution, we note a gradual increase of complexity of the interacting systems and related mechanisms involved: from ``simple'' chemical reactions between biomolecules, to life-sustaining circulation systems in plants, to the ultra-complex interaction chains in the human brain.

Any information-driven interaction between \textit{inanimate} complex systems must ultimately be life-generated or -designed, requiring at some stage goal-directed actions by a living system. Examples are the physical effects that a beaver dam has on water flow; mechanical effects on the environment of a tool used by a corvid; flight path control by an autopilot, etc. As a more explicit example, consider an electromagnetic or sound wave emitted by a meteorological lightening discharge, which does not represent any information-driven interaction: it is generated and propagates through physical processes in which information plays no role. But waves emitted by an electric discharge in the laboratory may be part of an overall artificial information-driven interaction mechanism if they are part of a device created by a human mind with the intention of having the discharge cause a desired change somewhere else, such as a record on a tape or a mental event in an observer. 

\section{Information and Classical Physics}

All information-driven interactions, whether purely biological or in a human-controlled scientific experiment, affect the ``normal'' non-biological course of physical and chemical events. Information-driven interactions all involve complex systems in the classical domain, with time-sequences which fulfill the dictates of causality, locality, special relativity and thermodynamics. It is important to emphasize that information-driven interactions indeed do function on the basis of force-driven interactions between their components---what makes them different is how these purely physical components \textit{are put together} in the interaction mechanism (the ``informational architecture'' of Walker et al., 2016). The important question of \textit{when} a force-driven interaction becomes information-driven will be briefly addressed later.  

The concept of ``information'' does not appear as an active, controlling agent in purely physical interaction processes in the Universe; it only appears there when a life system in general, or an \textit{observer} in particular, intervene (see Roederer 2005, Chapter 5). In other words, ``the world out there'' works without information-processing---until a living system intervenes and changes the physical course. The above-mentioned beaver dam is an example. In physics, when we state that ``a system of mass points follows a path of least action'', we do not mean that the system ``possesses the necessary information to choose a path of  minimum-action'' from among infinite possibilities, but that it is \textit{we} humans who have discovered \textit{how} systems of mass points evolve and who developed a mathematical method applicable to all to predict or retrodict their motions. Similar arguments can be made when we describe black holes as ``swallowing information'', or decoherence as ``carrying away information on a quantum system''. In summary, the physical universe does not ``obey'' physical laws---it is \textit{us the observers} who ``make'' the laws based on systematic observations (ultimately, specific changes in our brains---see section 3), of the changes left in our instruments by their exposure to environmental events. And it is us who are able to devise a mathematical framework that entices us to \textit{imagine} (see section 3) processes like ``black holes being an information-processing and -erasing system'', or ``an electron being at two different places at the same time''.     

Further examples are found in the association between entropy and information, which arises from the particular way \textit{we} scientists describe, analyze and manipulate nature, for instance, by counting molecules in a pre-parceled phase space; coarse-graining (averaging over preconceived domains); looking for regularities vs. disorder; quantifying fluctuations; extracting mechanical work based on observed patterns in the system; or mentally tagging molecules according to their initial states. In this latter context, let us discuss a concrete example, which also will be helpful later in a discussion of information in the quantum domain. We turn to Gibbs’ paradox (for details see Roederer (2005), section 5.5). Consider two vessels {A} and {B} of equal volume, joined by a tube with a closed valve, thermally isolated and filled with the same gas at the same pressure and temperature. If we open the valve, nothing will happen thermodynamically, but at the microscopic level, we can picture in our mind the molecules of {A} expanding into vessel {B} and the molecules of {B} expanding into {A}. Each process would represent an adiabatic expansion with an increase of the entropy by $-k N \ln{2}$ (where $k$ is Boltzmann’s constant and $N$ is the number of molecules in {A} or {B}), so there should be a total increase of the entropy of the system by ${\Delta S=-2 k N \ln{2}}$. This of course is absurd and represents Gibbs’ paradox. In most textbooks it is (somewhat lamely) explained away by saying that the formulas used here are indeed correct, but apply \textit{only} if the gases in the two vessels are  \textit{physically distinguishable}\footnote{Notice the link with the entropy increase/decrease $\epsilon$ per bit of Shannon information (see Roederer, 2005): ${\epsilon = {\Delta S}/(2N) =-k \ln{2}}$.} (different gases, or just specimens of opposite chirality). 

So what is wrong with the conclusions of the above thought experiment? By invoking the concept of pragmatic information when we say that ``we can picture in our mind the molecules from vessel {A} doing this or that…'', we are \textit{labeling} them so as to make them different from the others, even if in reality they are not. In other words, we are assuming that on each molecule there is a pre-established pattern (which triggers in our brain the neural correlates ``this molecule is from {A}, that molecule is from {B}''). However, this pattern and any extractable information from it \textit{do not exist}---we just have forced them into our mental image! The same argumentation, invoking the concept of pragmatic information where there is none in reality, can be made regarding Maxwell’s Demon paradox. Notice carefully that whenever we invoke pragmatic information in our mind, as in a Gedankenexperiment (i.e., establishing some imagined correlation between a pattern and a change governed by some imagined interaction mechanism), we are dealing with a thermodynamically \textit{open system}, even if in reality it is not.   

Notice that in all examples above, the interactions involved are force-driven; but whenever we use the term information in their description we really mean ``pragmatic information \textit{for us}, the observers''. And when \textit{it is us} who deliberately set the initial conditions of a classical mechanical system (or prepare a quantum system), we are converting it into an information-driven system with a given purpose (to achieve a change that would not happen naturally without our intervention). All laboratory experiments, whether a simple classroom demonstration or a sophisticated table-top quantum experiment, fall into this category. 

It will be useful for our discussion of quantum systems to identify common features in a classical measurement process. First of all, note that it necessarily involves force-driven interactions, but they are controlled by a human being or a preplanned artifact. For instance, when you measure the size of an object with a caliper, the instrument interacts elastically with the object; if you measure it with a ruler, you need to submerge everything in a ``bath of photons'' whose scattering or reflection is what your optical system uses to extract the wanted information on ``object--apparatus'' interaction. Note that in the latter example, \textit{the environment} has become an integral part of the instrument (this will be important for the discussion of quantum measurements). In summary, the measurement process represents an information-driven interaction between the object or system to be measured and a measurement apparatus or device (which contains a reference which we call ``the unit''). The whole process \textit{defines} the magnitude (the observable in quantum mechanics) that is being measured.\footnote{e.g., \textit{weight} of an object is the number we obtain when we do this and this and that, under these or those conditions. An important requirement in this is that different equivalent ``recipes'' yield the same number.} And it has an observer-related purpose: that of changing his/her state of knowledge in a very specific way. It is clear then, that even in classical physics, it is impossible to detach the measurement process from the observer (or his/her measurement artifacts). In summary, the fundamental interaction stages in any measurement process are: [patterns from an object] $\rightarrow$ [change in the apparatus], and [pattern of change in the apparatus] $\rightarrow$ [change in the state of the brain of the observer], who thus has \textit{extracted pragmatic information from the object}. Of course, in most cases that information can be represented as a number (value of an observable) and linked to Shannon or algorithmic information.

And here we should turn to another fundamental, inextricable link of the concept of information to biology, by pointing out that all the operations mentioned above are ultimately related to how the human brain of the observer creates internal mental images, plans, makes decisions, and reacts to and processes external sensory information.  As already mentioned above, these operations ultimately have the purpose of changing the neural cognitive state from ``not-knowing'' to ``knowing'', a transition which I venture to describe as the reduction of an initial brain state involving multiple expectancies to one of possible ``basis states'', where each basis state represents the mental image of only one possible outcome of the expected alternatives (we might even  call them ``preferred mental states''). Note that as a corollary, the concept of probability, usually defined mathematically as the result (limit) of a specific physical operation (e.g., a series of measurements under equal conditions like tossing of dice, playing roulette), has a very subjective foundation in human brain operation. 

\section{Information and Brain Function}

Since the beginning of quantum mechanics, physicists have been arguing about whether the observer and his/her state of knowledge, even consciousness, play an active role in the quantum measurement process. However, they did not have the benefit of knowing what is known today about the neurobiological mechanisms that control human brain function.

Recent studies with functional magnetic resonance imaging, positron emission tomography, diffusion tensor imaging and, at the neural network level, multi-microelectrode recordings, are confirming a hypothesis long in use by neurophysicists and computer scientists, namely that the information being processed in the brain is encoded in real-time as a task-specific \textit{spatio-temporal distribution of neural activity} in certain regions of the cerebral cortex (e.g., Tononi and Koch 2008). 
If, because of neural interconnectivity, a certain specific pattern of neural activity distribution in one area triggers a specific distribution in another, and does so in univocal way (within limits), we are in presence of an information-based interaction between two cerebral regions; the pragmatic information involved \textit{represents} that specific relationship, and we usually say that information has been transferred from one cerebral region to another. 

When a scientist makes a measurement, the pragmatic information involved represents the correlation between an external pattern (e.g., the location in a reference space of a rigid body at a given time, the position of the dial in an instrument, the dots on cast dice, the color change of a solution) with a specific spatio-temporal pattern of neural activity in the prefrontal lobes, corresponding to the knowledge ``it's \textit{this} particular state and not any other possible one''. The actual information-processing mechanisms in the brain linking one neural distribution with another are controlled by the actual synaptic wiring, which in certain regions, especially the hippocampus, has the ability of undergoing specific changes as a function of use (``plasticity")---the physiological expression of stably stored pragmatic information or long-term memory.

Modern neurobiology has an answer to the common question: When does a specific distribution of neural firings actually become a mental image? This neural activity distribution does not \textit{become} anything---it \textit{is} the image!\footnote{Just as the idea of information being a ``purely'' biological concept is unpalatable to many physicists, the idea of a mental image, even consciousness itself, being ``nothing but'' a very specific, unique spatio-temporal distribution of neural activity is unpalatable to many psychologists, philosophers and theologians.} In summary, the dynamic spatio-temporal distribution of neural impulses and the quasi-static spatial distribution of synapses and their efficiencies together are the physical realization of the global state of a functioning brain at any instant of time. Another way of expressing this: pragmatic information is encoded in the brain dynamically in short-term patterns of neural impulses and statically in the long-term patterns of synaptic architecture. Given the number of interacting elements ($\sim10^{12}$ neurons and $\sim10^{14}$--$\sim10^{15}$ synapses in the human brain) and the discontinuous nature of activity distribution, there is little hope that a quantitative mathematical theory of integral brain function would be developed in the foreseeable future.

Quite generally, animal brains handle pragmatic information in sequences of information-driven interactions in which one specific spatio-temporal pattern of neural activity is mapped or transformed into another neural pattern---in its most basic form, from a physically triggered sensory or interoceptive stimulation pattern to a neural output pattern controlling muscle and gland fibers, thus governing the animal's integral behavior. These processes may change the interconnectivity (synaptic architecture) of participating neural networks (the learning process), leading to long-term storage of information. A memory recall consists of the replay of the original neural activity distribution that had led to the synaptic changes during memory storage; the most important type is the \textit{associative recall}, in which the replay is triggered by a cue embedded in the ongoing neural activity distribution (for examples, see Roederer, 2005). Expressed in terms of pragmatic information: in the act of remembering or imagining a certain object, information on that object stored in the synaptic architecture of the brain is retro-transferred in the form of neural impulse patterns, mostly via subcortical networks, to the visual cortex and/or other pertinent sensory areas, where it triggers neural activity specific to the \textit{actual} sensory perception of the object. Neuroscientists call the specific microscopic distribution of neural activity responsible for any subjective experience a \textit{neural correlate}. 

Let us point out at this stage that in the definition of pragmatic information, it is often the case that different source patterns can lead to the same change in the recipient (e.g., different shapes, sizes and colors of an apple still trigger the neural pattern that defines the concept or image of ``apple''). Likewise, the same source pattern can lead to different effects, depending on collateral information-processing activity of the system. 

The human brain can recall stored information \textit{at will} as images or representations, manipulate them, discover overlooked correlations\footnote{In animals, the time interval within which causal correlations can be established (trace conditioning) is of the order of tens of seconds and decreases rapidly if other stimuli are present (e.g., Han et al. 2003); in humans it extends over the long-term past and the long-term future (for a brief review of human vs. subhuman intelligence, see for instance Balter 2010). Most importantly for our discussion, this leads to the conscious awareness of the past, present and future, and the quantitative conception of time.}, and re-store modified or amended versions thereof, \textit{without any concurrent external or somatic input}---it can go ``off line'' (Bickerton 1995). This is \textit{information generation} par excellence and represents the \textit{human thinking process} (e.g., Roederer 1978, 2005). Internally triggered human brain images, however abstract, are snippets (expressed as many different but unique patterns of neural activity in specific regions of the cortex) derived from stored information acquired in earlier sensory or mental events, and pieced together in different ways under some central control (the ``main program'') linked to human self-consciousness. 

Whenever a physicist conceives or thinks about the model of a physical system or physical process, whether classical, relativistic or quantum, whether one-dimensional or multi-dimensional, his/her brain triggers, transforms and mutually correlates very specific and unique distributions of neural impulses. The fact that the brain is an eminently classical information-processing device\footnote{Quantum decoherence times in the brain cells would be ten or more orders of magnitude shorter than the minimum time required for any cognitive operation (e.g., Schlosshauer 2008).} that evolved, and is continuously being trained through information-driven interactions with the classical macroscopic world, is very germane to how we can imagine, describe and understand the behavior of any systems, either quantum, classical, molecular or biological. This even applies to mathematics, most notably probability theory (see Roederer, 2005, section 1.6). For instance, given a set of mental images of possible outcomes of a future quantum measurement, all may be subjectively viewed as equiprobable by an unbiased observer. Only after personal experience with multiple measurements under identical conditions, or through information from others who already have undertaken this task, can the observer develop an objective sense of traditional probability.

\section{Information and Quantum Mechanics:\\ ``Bits'' from ``Its''}

In the preceding sections we have identified two categories of interactions between bodies or systems in the universe as we know it: force-driven and information-driven. The first category is assumed to be operating in the entire spatial-temporal domain, from the Planck scale up, between elementary particles, atoms, molecules, complex condensed bodies, networks and systems of bodies. The second category leads to the definition of pragmatic information as representing a physical, causal and univocal correspondence between a pattern and a specific macroscopic change elicited by some complex interaction mechanism. By this very definition, the domain of validity of information-driven interactions, and therefore of the concept of pragmatic information per se, is limited to the classical macroscopic domain. The reason is that, as we shall discuss extensively below, in the subatomic quantum domain spatial or temporal patterns cannot always be defined or identified in a univocal, stable or causal way. This in turn is related to fundamental and exclusive properties of quantum systems, which we shall examine now taking into account the definition of pragmatic information given in section 1.

\subsection{Revisiting some relevant quantum facts}

For the purpose of exploring the role of pragmatic information in quantum mechanics it is opportune to present, in greatly simplified fashion, some experiment-based ``quantum facts in a nutshell’’.\footnote{An excellent discussion of the principles of quantum mechanics as relevant to this article is given in Chapter 2 of Schlosshauer (2008).} To emphasize that the peculiar behavior of a quantum system is not the result of some mathematical properties of linear algebra and differential equations but, rather, representative of \textit{physical} facts happening ``out there'', we shall deliberately refrain from referring to abstract postulates, theorems and properties of Hilbert spaces with which most textbooks  introduce the subject. 

Given a \textit{single} quantum system that has been physically prepared in a certain way, it is \textit{impossible in principle} to determine or verify its particular state by a measurement. Indeed, having many similar quantum systems prepared in identical ways and subjecting each one to identical measurements, one will obtain a collection of measurement values from among a common (often discrete) set that only depends on the setup and the instrument, called the \textit{eigenvalues} of the observable in question (as mentioned in the previous section, the observable itself being defined by the instrument). The distribution of \textit{probabilities of occurrence} of eigenvalues (including the latter's mean value and standard deviation) is what characterizes the common state of each quantum system of the set at the time of measurement (Born's Rule). While it is impossible to predict the outcome of the measurement of any \textit{one} system, the set of probabilities of obtaining the eigenvalues is deterministic in the sense that it depends only on the preparation and history of the quantum system and on the apparatus used.\footnote{There are classical systems that do behave this way---for instance a pinball machine or a roulette. They are the so-called deterministic chaotic systems, which \textit{during the preparation process} suffer unpredictable and unavoidable infinitesimal changes which then lead to \textit{finite} differences with which the system comes out of the preparation process. Table-top experiments with quantum systems involving single particles (e.g., Mach-Zehnder interferometry, electron diffraction) convincingly show that no such \textit{final} differentiation happens during preparation.} This is not all: when a single quantum system is measured and one of the possible eigenvalues of the corresponding observable is obtained, that quantum system emerges in a special state such that a repeated measurement of the exactly same kind immediately thereafter will always yield the same eigenvalue.\footnote{We exclude so-called destructive measurements, in which the quantum system disappears (e.g., is absorbed).} Such a state is therefore called an \textit{eigenstate} (also ``basis state"). One says that the initial state of the quantum system has \textit{collapsed} into, or been \textit{reduced} to, an eigenstate as the result of the first intervention. In other words, after the measurement the quantum system behaves ``classically'' \textit{with respect to the variable in question}.

Sometimes, the same apparatus can be used in different configurations (e.g., rotation of a Stern-Gerlach set-up or a polarization filter); experiments show that each configuration may have a different set of basis states. In other cases there are no alternatives for the same instrumental setup (for instance, the ``which-path'' observable in a Mach-Zehnder interferometer; the wavelength of a spectral emission; the decay products of a particle); the eigenstates are then called ``preferred basis states''. On the other hand, one single quantum system may have more than one observable to measure (several degrees of freedom like the position, momentum and spin vectors of a particle); in that case, the measurement results of one observable may or may not affect (be correlated with) the distribution of measurement results of another. This means that having many multi-degree of freedom quantum systems prepared in identical ways, making the measurement of one specific observable on some and measurement of another observable on the rest, the statistical distributions of results in both series of measurements may or may not be correlated (non-commuting or commuting variables, respectively).

It is when non-commuting observables enter the picture that Planck's universal constant $\hbar$ appears and delimits the ``size'' of the quantum domain. Historically, it was the position and the momentum of a particles which led Heisenberg to his uncertainty relationship between their standard deviations in measurements under similar conditions ($\Delta x \Delta p \ge {\hbar} /2$). For the sake of completeness of this summary, we must also mention that Schr\"{o}dinger's equation is introduced when the time evolution of a quantum system is to be described (which, however, will not be dealt with in this article).

All these experimental facts indeed allow the development of linear algebra algorithms in which general states of a quantum system (those which, when measured under equal conditions, give different results from among a set of eigenvalues) are represented by (unit) vectors in a Hilbert space; interactions such as measurements and transformations (e.g., mutual interactions, time evolution) are represented by specific operators. Each observable defines a subspace whose axes represent its eigenstates or basis states when measured with the observable-defining apparatus in a specific configuration, and the squares of the projections of the state vector onto the axes represent the probabilities of occurrence of the corresponding eigenvalues. A general quantum state (called a pure or superposed state) is thus represented as a \textit{linear superposition of basis states}; a measurement is represented by the rotation of the state vector onto one of the axes.\footnote{Complex numbers are used for the state vector components to accommodate the possibility of having two or more superposed states with the same probability distribution of measurement results, but which behave differently in interactions (interference phenomena).} Finally, the functional expressions (e.g., the Hamiltonians) governing the dynamic evolution of the obervables of a quantum system are equivalent, in many cases identical, to the corresponding functional expressions for an equivalent classical system. 

Here comes our first real encounter with the concept of information. The mathematical formalism developed for quantum systems in correspondence with the formalism of classical mechanics tempts us to picture \textit{in our mind} a superposed state as the quantum system ``being in different eigenstates at the same time''. As stated before, for a single quantum system it is impossible to determine through measurement whether it is or not in such a superposition, and if it is, what that superposition really is. In other words: it is impossible to extract pragmatic information from a single quantum system. This is no different than wanting to imagine in the above-mentioned Gibbs' paradox the molecules from one container moving into the other preserving a distinguishability as if they were carrying labels from which to extract pragmatic information---when in reality there is none. 

Let us consider an ensemble of $N$ equally prepared, identical quantum systems. We measure one given variable on each one and compile the results, finding a whole set of different outcomes in a certain statistical distribution. This tells us that the \textit{preparation} process had produced an ensemble of quantum systems, each one in the same \textit{coherent superposed state}. Now \textit{after} the collective measurement was made, we are left with a radically different kind of ensemble, namely a \textit{mixture} of $N$ systems in which each component is now collapsed into a fixed, repetitively measurable eigenstate.\footnote{Remember that in all this we are ignoring possible changes as a function of time.} Pulling out just one component at random, we would never be able to predict what its state is. But after the initial measurements we could have \textit{labeled} each element with the pertinent result (knowing that a repeated measurement would yield the same value) and obtain a mixture of systems that would behave classically with respect to the observable in question, because now we \textit{can} extract pragmatic information from it (e.g., we could create stable patterns and establish univocal correspondences with changes or new patterns elsewhere).

Next, we consider an ensemble of several distinct but initially \textit{non-interacting} quantum systems in superposed states. Each one is described by its own separate state vector in its own Hilbert space (corresponding to the observable(s) chosen). If measured individually, we obtain individual distributions of results for each component. Let us now, instead, bring these individual quantum systems into mutual interaction at the quantum level (i.e., shielded from interactions with macroscopic bodies), take them away from each other, and only then make the measurements. Like before, we cannot predict what the individual outcomes will be, but once obtained, experiment tells us that they appear \textit{correlated}--regardless how far apart they were taken. This means that once a set of quantum systems in superposed states interact, they will lose their independence: the ensemble must be treated as \textit{a single whole}, described by just \textit{one} global state. These component parts are then said to be \textit{entangled}. The actual results of measurements on entangled components (unpredictable in themselves, except statistically) will be correlated, no matter \textit{how far away in space and time} they are located from each other after the mutual interaction (there is no equivalent example of this in classical physics). It is often stated: ``once the measurement has been made on one system, the other one will automatically collapse into a conjugate basis state’’. Yet there is no information transfer involved in this process, no superluminous communication between the two! All we find (at the macroscopic level) is that the measurement results will be correlated--but correlation does not imply causation!

It is also an experimental fact that not only man-made laboratory measurements but \textit{any} interaction with the natural macroscopic environment will eventually break up the global state of an ensemble of entangled quantum systems into independent, ultimately collapsed states of its individual components. This process is called \textit{decoherence} (e.g., Schlosshauer, 2008). As a matter of fact, it is impossible to completely shield a quantum system from unpredictable macroscopic influences of the environment, and superposed quantum states are difficult to maintain in the laboratory.

Entanglement allows us to take a better look at the quantum measurement process, in which a quantum system is deliberately made to interact with a macroscopic device. This device is constructed in such a way that in the special case when the quantum system is in an eigenstate of the observable which the apparatus is supposed to measure, the initial perturbation propagates through the instrument and is amplified to give an observable macroscopic effect (e.g., the particular position of a dial) that depends on the initial eigenstate. When the quantum system to be measured is now in a superposed state, it gets entangled with the ``quantum end’’ of the apparatus---but as subsequent interactions propagate through the instrument, inevitable entanglement with the environment will eventually cause the single quantum state of the total system (measured system plus instrument) to break up into mutually correlated basis states: (i) the instrument signals a specific macroscopic change, and (ii) the original quantum system emerges in a corresponding eigenstate.\footnote{The ``Schr\"odinger Cat'' paradox refers to what would happen if there was no decoherence in this process.} Because of the human intention involved in the construction or use of a measurement apparatus, does a quantum measurement process represent an information-driven interaction? There was no initial pattern to be univocally mapped into a macroscopic change (like the position of a pointer)---the only information-driven interaction occurs between the ``classical end'' of the instrument and the observer (sensory effect of the macroscopic change in the instrument on the observer's brain or a human-designed recording device). 

When an ensemble of quantum systems has decohered, it will behave classically, at least with respect to the observables whose states have collapsed, furnishing the \textit{same} results in immediately successive, identical measurements. Only then can we define patterns based on measurement outcomes for information-driven interactions---such interactions cannot occur at the non-decohered quantum level. We can emphasize again: \textit{pragmatic information cannot exist and operate in, or be extracted from, a pure quantum domain}. A logical consequence is that a single quantum system in a superposed state cannot be copied (the so-called no-cloning theorem). We'll come to this again in the next section. 

It is an experimental fact that macroscopic bodies are in decohered and reduced states, particularly of the position variable, but so are also the atomic nuclei of organic molecules.\footnote{But not their subatomic constituents (electrons, nucleons, quarks). This does not mean that under very special laboratory conditions entire molecules cannot find themselves as a whole in a superposed quantum state---diffraction experiments with $C^{60}$ molecules have been carried out successfully (Hackerm\"{u}ller et al. 2004).} The latter behave classically and can carry, transfer or respond to pragmatic information: a spatial pattern in some organic molecule (e.g., RNA) triggering a specific change in some other molecular system (e.g., formation of a protein), with some complex molecular organelle (e.g., a ribosome) responsible for the interaction mechanism, is possible! Herein lies the essence of the evolutionary emergence of life systems (Roederer, 2005).

Let us now walk through some well-known simple examples, usually taught at the beginning of a course on quantum mechanics or quantum computing, and examine them from the point of view of pragmatic information---pointing out where the latter plays a role and where it can't.

\subsection{\textbf{Single qubits}}

Consider the measurement of a qubit (elementary quantum system with only two possible eigenstates, like a 1/2 spin silver atom, a photon in a two-path Mach-Zehnder interferometer). The qubit interacts with the apparatus at its quantum end (e.g., the atoms to be ionized in a particle detector), and after the measurement the classical end exhibits a macroscopic change (e.g., voltage pulse, a blip on a luminescent screen, position of a pointer, a living or dead cat). It is the physical structure of the device enabling the occurrence of such macroscopic change that \textit{defines} the observable in question (although in this case it is better to say that, given the qubit, its two preferred basis states determine the instrument that must be used). The observer per se is irrelevant during the measurement itself, except that we must not forget that it was a human being who set up the measurement (hence decided what observable to measure), who selected and prepared the qubit to be measured, and whose brain ultimately expects to receive an image, \textit{the neural correlate}, of the change of the macroscopic state of the apparatus as a result of the measurement (knowledge of the value of the binary observable). 

Following the usual von Neumann protocol, let us call $\left|\textit{M}\right\rangle$ the initial state of the apparatus and 
$\left|\textit{M}_{0}\right\rangle$,$\left|\textit{M}_{1}\right\rangle$ 
the two possible alternative states of the apparatus \textit{after} the measurement. The instrument is \textit{deliberately} built in such a way that when the qubit to be measured is in basis state $\left|0\right\rangle$  before its interaction with the apparatus, the final independent state of the instrument after the measurement will be $\left|\textit{M}_{0}\right\rangle$, and if the state of the qubit is $\left|1\right\rangle$  the instrument will end up in state $\left|\textit{M}_{1}\right\rangle$. In either case, the state of the qubit remains unchanged (we are assuming this to be a non-destructive measurement). Therefore, for the \textit{composite} state qubit--apparatus we will have the following evolution in time, as determined by the Schr\"{o}dinger equation: 
$\left|0\right\rangle\left|\textit{M}\right\rangle\rightarrow\left|0\right\rangle\left|\textit{M}_{0}\right\rangle$ \textit{or} $\left|1\right\rangle\left|\textit{M}\right\rangle\rightarrow\left|1\right\rangle\left|\textit{M}_{1}\right\rangle$.

If the qubit is now in a \textit{superposed state} $\alpha|0\rangle+\beta|1\rangle$ (with $\left|\alpha\right|^2$+$\left|\beta\right|^2$=1), since the Schr\"{o}dinger equation is first-order in time we will obtain an \textit{entangled} state and the state of the composite system qubit--apparatus will remain a linear superposition, as long as it is kept isolated from all other interactions: $(\alpha|0\rangle+\beta|1\rangle)|\textit{M}\rangle\rightarrow\alpha|0\rangle|\textit{M}_{0}\rangle+\beta|1\rangle|\textit{M}_{1}\rangle$. However, it is an experimental fact that one has never observed a macroscopic system with such peculiar properties as superposition (the essence of the Schr\"{o}dinger cat paradox): the end state of the composite system will always be either 
$|0\rangle|\textit{M}_{0}\rangle$ or $|1\rangle|\textit{M}_{1}\rangle$ (decoherence) with $\left|\alpha\right|^2$ and $\left|\beta\right|^2$ the probabilities to obtain either result, respectively, in a large set of measurements under strictly identical conditions from preparation to end result (Born rule). In each process, the instrument comes out in the macroscopic state $|\textit{M}_{0}\rangle$ or $|\textit{M}_{1}\rangle$, and the  original qubit emerges in the corresponding eigenstate (state reduction). 

According to our definition of pragmatic information, decoherence and state reduction thus express the fundamental fact that \textit{no information can be extracted experimentally on the superposed state of a single qubit}. A direct consequence of this is the fact, already mentioned before, that the state of one given qubit cannot be copied. Indeed, if we \textit{could} make \textit{N} ($\rightarrow\infty$) copies of a single qubit in a superposed state, a correspondence could indeed be established between the original pair $\alpha$, $\beta$ and some macroscopic feature linked to the statistical outcome of measurements on the \textit{N} copies (in the case of a qubit, for instance the average values of some appropriate observables). This would be tantamount to extracting pragmatic information from the original \textit{single} qubit. What is possible, though, is to repeat exactly \textit{N} times the \textit{preparation process} to obtain \textit{N} separate qubits in the same superposed state (like retyping a text on blank sheets repeatedly, instead of Xeroxing the original \textit{N} times). Each qubit of this set can then be subjected to a measurement, and the parameters 
$\alpha,\beta$ extracted from the collection of results (this is, precisely, how the probabilities $\left|\alpha\right|^2$ and $\left|\beta\right|^2$ are obtained experimentally\footnote{Remember that since $\alpha$ and $\beta$ are two normalized \textit{complex} numbers, in order to determine, say, their relative amplitudes and phase statistically, it is necessary to obtain \textit{two} sets of measurement data (\textit{N}/2 measurements in each set).}).

The \textit{preparation} of a qubit in a given superposed state requires the intervention of complex macroscopic devices and three steps of action on a \textit{set} of qubits: 1) measurement of an appropriate observable (which leaves each qubit in an eigenstate of that observable); 2) selection (filtering) of the qubits that are in the desired eigenstate (a classical process); 3) unitary transformation (a rotation in Hilbert space) to place the selected qubit in the desired superposed state. In this procedure a preparer has converted a certain \textit{macroscopic} pattern (embedded in the physical configurations of the preparation process) into the values of two complex parameters of a quantum system (e.g., the normalized $\alpha$ and $\beta$ coefficients in the mathematical description of the qubit). According to our definition, doesn't such correspondence represent genuine pragmatic information (correspondence pattern $\rightarrow$ change)? No, because it would not be univocal: given a \textit{single} qubit in an unknown state, an observer could never reconstruct through measurement the original macroscopic configuration used, or steps taken, in the preparation process. The only way to do so is to remain in the macroscopic domain and \textit{ask} the preparer (classical information from brain to brain!). Once known via such a macroscopic route, it will in principle be possible to \textit{verify} (but not to determine from scratch) the information about $\alpha$ and $\beta$.
 
\subsection{Entangled qubits and space-time}

Take two qubits \textit{A} and \textit{B} that are maximally entangled in the antisymmetric Bell state 
$\Psi^- = 1/\sqrt{2}(|0\rangle_{A}|1\rangle_{B}-|1\rangle_{A}|0\rangle_{B})$ 
at time $\textit{t}_{0}$. We may imagine qubit \textit{B} now being taken far away. If nothing else is done to either, we can bring \textit{B} back, and with some suitable experiment (e.g., interference) demonstrate that the total state of the system had remained entangled all the time. If, instead, at time $\textit{t}_{A}>\textit{t}_{0}$ a measurement is made on qubit \textit{A} leaving the composite system reduced to either state $|0\rangle_{A}|1\rangle_{B}$ \textit{or} $|1\rangle_{A}|0\rangle_{B}$, qubit \textit{B} will appear in either basis state $|1\rangle_{B}$ \textit{or} $|0\rangle_{B}$, respectively, if measured. The puzzling thing is that it does not matter \textit{when} that measurement on \textit{B} is made---even if made \textit{before} $\textit{t}_{A}$. Of course, we cannot predict which of the two alternatives will result; all we can affirm is that the measurement results on each qubit will appear \textit{to be correlated}, no matter the mutual spatial distance and the temporal order in which they were made\footnote{One could argue that in the case of a measurement on \textit{B} at an earlier time $\textit{t}_{B}<\textit{t}_{A}$, it was \textit{this} measurement that ``caused'' the reduction of the qubits' composite state--but the concepts of ``earlier'' and ``later'' between distant events are not relativistically invariant properties. See also next paragraph.}  

The result of all this is that it appears as if the reduction of the quantum state of an entangled system triggered by the measurement of one of its components is ``non-local in space and time''. Yet as stated before, correlation does not mean causation in the quantum domain: nothing strange happens at the macroscopic level: the state reduction cannot be used to transmit any real information from \textit{A} to \textit{B}. In terms of our definition of pragmatic information, there is no ``spooky'' action-at-a-distance: an experimenter manipulating \textit{A} has no control whatsoever over \textit{which macroscopic change} shall occur in the apparatus at \textit{B}, and vice versa. The spookiness only appears when, in the \textit{mental} image of a pair of spatially separated entangled qubits, we force our (macroscopic) concept of information into the quantum domain of the composite system and think of the act of measurement of one of the qubits as \textit{causing} the particular outcome of the measurement on the other.

Yet another insight can be gleaned from the re-examination of the so-called \textit{quantum teleportation} of a qubit (e.g., Bouwmeester et al., 1997). Let us remember the basic procedure: an entangled pair of qubits in the antisymmetric Bell state 
$\Psi^- = 1/\sqrt{2}(|0\rangle_{A} |1\rangle_{B} -|1\rangle_{A} |0\rangle_{B})$
is produced at time $\textit{t}_0$  and its components are taken far away from each other. 
At time $\textit{t}_{A} >\textit{t}_{0}$ an unknown qubit in superposed state 
$\alpha|0\rangle_{C} + \beta|1\rangle_{C}$ 
is brought in and put in interaction with qubit \textit{A}. The total, composite, state of the three-qubit system is now
$1/\sqrt{2}(\alpha|0\rangle_{C}+\beta|1\rangle_{C}) (|0\rangle_{A}|1\rangle_{B}-|1\rangle_{A}|0\rangle_{B})$, 
which can be shown algebraically to be equal to a linear superposition of four Bell states in the \textit{A-C} subspace, with coefficients that are specific unitary transforms of the type 
${(-\alpha|0\rangle_{B}-\beta|1\rangle_{B})}$, ${(+\alpha|1\rangle_{B}-\beta|0\rangle_{B})}$, and so on. Therefore, if a measurement is made on the pair $\textit{A-C}$ of any observable whose eigenstates are the four Bell states, the state of the entire system will collapse into just one of the four terms, with the qubit at \textit{B} left in a superposed state with coefficients given by the parameters $\alpha, \beta$ of the now vanished unknown qubit \textit{C}. If the observer at point \textit{A} informs \textit{B} (a classical, macroscopic transfer of pragmatic information) which basis Bell state has resulted in the measurement---only two bits are needed to label each possible basis state---observer \textit{B} can apply the appropriate inverse unitary transformation to his qubit, and thus be in possession of the teleported qubit \textit{C} (defined by the unknown coefficients $\alpha$ and $\beta$).

The puzzling aspect of this procedure is that it looks as if the infinite amount of information on two real numbers (those defining the normalized pair of complex numbers $\alpha$ and $\beta$) was transported from \textit{A} to \textit{B} by means of only two classical bits. The answer is that, again, according to our definition, $\alpha, \beta$ \textit{do not represent pragmatic information} on the state of any \textit{given} qubit. They are quantitative parameters in the mathematical framework developed to describe quantum systems and their interactions, but they cannot be determined physically ``out there'' for a given qubit. Related to this, there is no way to verify the teleportation of a \textit{single} qubit; the only way verification could be accomplished is through a \textit{statistical} process, repeating the whole procedure \textit{N} times, from the identical preparation of each one of the three qubits \textit{A}, \textit{B}, \textit{C} to the actual measurement of the teleported qubit. If we determine the frequencies of occurrence $\textit{N}_{0}$ and $\textit{N}_{1} (= \textit{N}-\textit{N}_{0})$ of the 
$|0\rangle_{B}$ and $|1\rangle_{B}$
states, and express the teleported qubit state in its polar (Bloch sphere) form 
$|\Psi\rangle = \cos(\theta/2)|0\rangle + \exp(i\phi)\sin(\theta/2)|1\rangle$, it can be shown that the number of \textit{statistically significant} figures (in base 2) of $\theta$ and $\phi$ to be obtained is equal to the \textit{total number} \textit{2N} of bits transmitted classically (i.e., macroscopically) from $\textit{A}$ to $\textit{B}$, so that from the \textit{statistical} point of view, there is no puzzle at all! Moreover, this shows that there is \textit{no way} of teleporting pragmatic information and, as a consequence, macroscopic objects!

The preceding discussion says something about how we tend to think intuitively of time and space at the quantum level. A point in the 4-dimensional continuum of space-time is a mathematical abstraction, useful in the description of objects and events in the Universe. But \textit{position} and \textit{time intervals} of objects must be determined by measurements, i.e., with a macroscopic instrument, in which a macroscopic change is registered (e.g., a change in the ``bath of light'' in a position measurement; a change in the configuration of a clock or the angular position of a star). Like information, \textit{time is a macroscopic concept} (even an atomic clock must have classical components to serve as a timepiece). We can assign time marks to a quantum system only when it interacts \textit{locally} with (or is prepared by) a macroscopic system. In the case of a wave function $\Psi(\bf{x},\textit{t})$, the time variable refers to the time, measured by a macroscopic clock \textit{external} to the quantum system, at which $|\Psi|^2$ is the probability density of actually observing a quantum system at the position \textbf{x} in configuration space, which is also based on a measurement with a macroscopic instrument. 

Non-locality in space and time really means that for a composite quantum system, the concepts of distance and time interval between different superposed or entangled components \textit{are undefined} as long as they remain unobserved, i.e., free of interactions with macroscopic systems\footnote{Already in 1927 Heisenberg declared: “A particle trajectory is created only by the
act of observing it”!} (for a postulate on atemporal evolution, see Steane, 2007). Because of this, it may be unproductive trying to find a modified form of the Schr\"{o}dinger equation, or any other formalism, to describe quantitatively what happens ``inside'' a quantum system during the process of state reduction.\footnote{For instance, a theoretical or experimental derivation of ``average trajectories'' in a double-slit experiment (e.g., Kocsis et al. 2011) provides the geometric visualization of something on which, for an \textit{individual} particle, pragmatic information could never be obtained!} But, finally, what about the decay process of an unstable particle (or nucleus, for that matter), which run on the proper time of the particle, as demonstrated long ago with the decay times of cosmic ray $mu$-mesons when observed from a reference frame fixed to Earth? As we shall briefly mention, decay processes may be linked to decoherence; if that is indeed the case, they would be controlled by interactions with the macroscopic environment \textit{as experienced by the quantum system} (reduction to decay products).  

\subsection{The process of decoherence}

Returning to decoherence, let me briefly address the still contentious question of what, if any, \textit{physical} processes are responsible for the transition from the quantum end of a measurement apparatus to its classical, observable one (for details, see, e.g., Schlosshauer 2008). For this purpose, let us consider a super-simplified \textit{model} of a measuring apparatus that consists entirely of mutually interacting qubits---lots of them, perhaps $10^{22}$ or $10^{23}$---with one of which the external qubit to be measured enters into unitary non-destructive interaction at time $\textit{t}_{0}$. In our model, the apparatus qubits represent a complex web in some initial or ``ready'' state, designed in such a way that, as the local unitary interaction processes multiply and propagate, only two distinguishable macroscopic end states $\textit{M}_{0}$ and $\textit{M}_{1}$ can be attained, realized as two macroscopic spatially or temporally different forms or patterns (the so called “pointer states”, represented in mutually orthogonal Hilbert subspaces of enormous dimensions). The key physical property of this construction is that for a qubit in a basis state the instrument's final macroscopic configuration will depend on the \textit{actual} basis state of the measured qubit (see discussion in subsection 4.2).   

If the qubit to be measured is now in a superposed state, the first physical interaction at time $\textit{t}_{0}$ would create an entangled state of the system ``qubit--first apparatus quantum element'' which through further unitary inter-component interactions would then expand to the entire composite system ``qubit--apparatus'' in a cascade of interactions and further entanglements throughout the apparatus. Accordingly, the classical end of the apparatus also should end up in a superposed state (the Schr\"{o}dinger cat!) and since in principle the interactions are unitary, the whole process would be reversible. Obviously, somewhere in the cascade of interactions there must be an irreversible breakdown of the entanglement between the original qubit and the instrument, both of which will emerge from the process in separate but correlated states, either $|0\rangle$  and $\textit{M}_{0}$, or $|1\rangle$  and $\textit{M}_{1}$, respectively. Extractable pragmatic information appears only at the classical end of this cascade. All this means that if exactly the same kind of measurement is repeated immediately on the qubit (assuming that it was not destroyed in the measurement process), one will obtain a result that \textit{with certainty} is identical to that of the original measurement.  

Measurement processes and their apparatuses are \textit{artifices}---human planned and designed for a specific purpose. However, note that the preceding discussion on decoherence can be applied to the case in which we replace the artificial measurement apparatus with the \textit{natural environment per se}. Just replace the word ``apparatus'' with ``the environment'' with which a given quantum system willy-nilly interacts and gets entangled. As long as this entanglement persists, the given quantum system will have lost its original separate state, and only the \textit{composite} quantum system--environment will have a defined state, however complicated and delocalized. Now, if the interaction with the environment leads to a \textit{macroscopic} change somewhere (potentially verifiable through classical information-extraction by an observer, but independently of whether such verification actually is made), it will mean that decoherence has taken place and that the state vector of the original quantum system will have been reduced to one of its original eigenstates pertinent to the particular interaction process. 

This is usually described as ``entanglement with the environment carrying away information on a quantum system'', or ``information about the system's state becoming encoded in the environment''. However, I would like to caution about the use of the term information in this particular context: there is no loss of pragmatic information in natural decoherence, because there wasn't any there in the first place! A much less subjective way is to say: \textit{``A quantum system continuously and subtly interacts with its environment and gets entangled with it; if decoherence occurs, a macroscopic change in the state of the environment will appear somewhere (information about which could eventually be extracted by an observer), and the state of the quantum system will appear reduced to some specific basis state in correspondence with the environmental change in question.''} Since this basis state is in principle knowable, the decohered qubit now belongs to the classical domain. Measurement instruments are environmental devices specifically designed to precipitate decoherence and steer it to into certain sets of possible final macroscopic states (preferred states, if no alternative sets exist). 

An ensemble of identically prepared quantum systems (e.g., a chunk of a recently separated, chemically pure radioisotope) thus turns probabilistic because it is unavoidably ``submerged'' in a gravitating, fluctuating, thermodynamic macro-world, and will decay into a mixture of quantum entities in eigenstates \footnote{With the decay times mainly depending on the original wave function of the individual nuclei, but also slightly perturbed by subtle but unavoidable interactions of the latter with the environment (leading to fluctuations in the exponential decay of the ensemble.)}  (e.g., with an $\alpha$-particle either still inside or already outside a nucleus). On the other hand, the laboratory measurement of a quantum system may be viewed as a case in which the environment was deliberately altered by interposing a human-made apparatus, which then altered in a ``not-so-subtle'' way the time evolution of the system (we should really say: the time evolution of potential macroscopic effects of the system) by greatly increasing the chance of decoherence. In summary, what we have called the cascade of entanglements in a quantum measurement also involves a stochastic ensemble of outer environmental components with which the instrument's components are in subtle but unavoidable interaction.

A collateral consequence of natural decoherence is that any peculiar quantum property like superposition will have little chance of spreading over a major part of a macroscopic object, which indeed will behave classically whenever observed---there always seems to be a natural limit to the complexity of a quantum system in a pure superposed state beyond which it will decohere. In other words, the classical macroscopic domain, in which life systems operate and information can be defined objectively, consists of objects whose constituents have decohered into eigenstates (mainly, of their Hamiltonians). Quantum behavior of a macroscopic system is \textit{not}  forbidden (a Schr\"{o}dinger cat \textit{could} be in a superposed state of dead and alive at the same time!), but its probability and duration would be ridiculously small. This also explains the fact that, as mentioned before, many artificial quantum systems are very unstable in a superposed state, and thus very difficult to handle in the laboratory---a fact that represents one of the biggest challenges to quantum computing. 

Finally, we should view all dynamics equations in quantum mechanics like the Schr\"{o}dinger equation as the tools for providing information on potential \textit{macroscopic} effects of a quantum system on the environment (or a measurement apparatus) under given circumstances, rather than describing the evolution of the quantum system per se. The ``amazing aspect'' of quantum mechanics is not its puzzling paradoxes, but the fact that a mathematical framework \textit{could} be developed that can be used successfully to determine statistical, objective probabilities for \textit{observable} macroscopic outcomes of interaction processes both, in artificial experiments and in the natural environment.

\section{Concluding Remarks: Quantum Pedagogy}

From the previous discussion it is advisable to refrain from using the classical concept of pragmatic information indiscriminately to represent mentally a quantum system---be it by thinking about it, mathematically representing it or manipulating it in Gedankenexperiments. Yet quite commonly we do, especially when we teach---but then, as mentioned above, we should not be surprised that by \textit{forcing} the concept of information into the quantum domain, mental images are triggered of ``weird'' behavior that is contradictory to our every-day macroscopic experience. 

To me, the problem of the interpretation of quantum mechanics is not just one of a philosophical nature but one of eminently pedagogical nature. For instance, how should one answer correctly the often-asked question: Why is it not possible, even in principle, to extract information on the actual state of a \textit{single}  qubit? Because \textit{by the definition of information}, to make that possible there would have to exist some physical paradigm by means of which a change is produced somewhere in the \textit{macroscopic} classical domain that is in one-to-one correspondence with the qubit's parameters immediately prior to that process. Only for eigenstates (basis states) can this happen---decoherence prevents the formation of any macroscopic trace of superposed states. In the case of an initially superposed state, the end state of the qubit will always appear correlated with the end state of the macroscopic system, i.e., will emerge reduced to a correlated or preferred basis state. In somewhat trivial summary terms, quantum mechanics can only provide real information on natural or deliberate \textit{macroscopic imprints} left by a given quantum system that has undergone a given preparation, eventually interact unitarily (reversibly) with other quantum systems forming a composite quantum system, which as a single whole interacts irreversibly with the surrounding macroscopic world.

So what are the coefficients in a qubit state like $\alpha|0\rangle+\beta|1\rangle$ or $|\Psi\rangle = \cos(\theta/2)|0\rangle + \exp(i\phi)\sin(\theta/2)|1\rangle$? They are \textit{parameters in a model representation} of the system in complex Hilbert space, which within an appropriate mathematical framework enables us to make quantitative, albeit \textit{only probabilistic}, predictions about the system's possible macroscopic imprints on the classical domain. We may call $\alpha$ and $\beta$ or $\theta$ and $\phi$ ``information'', and we do, based on the fact that we can prepare a quantum system in a chosen superposed state---the common usage of the terms ``quantum bit'' and ``quantum information'' testifies to this. Yet for a single qubit we cannot retrieve, copy or verify the numbers involved, which means we cannot establish a univocal correlation between the state of the qubit and any macroscopic feature. In other words, those parameters \textit{are not pragmatic information} (the only exception is when the qubit is in one of its basis states). This is why when we do call $\alpha$, $\beta$, $\theta$  and $\phi$  “information”, we are always obliged to point out its “hidden nature”! And in teaching, we always would have to mumble something about super-luminal speed of information, teleportation of real things, a particle being in different positions at the same time, etc. to satisfy our (classical world) imagination. In Richard Feynman's words, we always would have to emphasize that ...\textit{the} [quantum] \textit{``paradox'' is only a conflict between reality and your feeling of what reality ``ought to be''}. The whole framework of quantum information theory and computing is based on the consistency of this kind of \textit{classical} correspondence: the relation of a given initial set of qubits in prepared eigenstates (a classical input pattern) correlated through intermediate unitary quantum interactions in an appropriately shielded quantum computing device with another final set of qubits in basis states (a classical output pattern). This input-output correlation is what really should called \textit{quantum information}, a genuine category of pragmatic information. 

Obviously, during the time interval between input and output, any extraneous non-unitary intervention, whether artificial (a measurement) or natural (decoherence), will change or destroy the macroscopic input-output correlation. Indeed, in this interim interval, the proverbial mandate of ``don't ask, don't tell'' applies (Roederer 2005)---not because we don't know \textit{how}  to extract relevant information to answer our questions, but because pragmatic \textit{information per se does not operate in the quantum domain.}

Let me end by stating a personal opinion as a former physics teacher. When it comes to evaluating, or to teaching about, the ``realities out there'', it is the physicists who should ``get real'' and recognize the fact that the World, both physical and biological, does not operate on the basis of what happens in Mach-Zehnder interferometers, Stern-Gerlach experiments, two-slit diffraction laboratory setups, qubit teleportation and all these marvellous experiments designed and performed by humans. In reality, all such experiments, while providing answers to the inborn human inquiry about how our environment works, are but artificial intrusions poking into a Universe that does not care about linear algebra, Hamiltonians, and about information per se. These experiments and the ensuing human understanding were made possible only because of the emergence, at least on Planet Earth, of interactions based not on force fields alone, but on the evolution of ultra-complex macroscopic mechanisms responding exclusively to simple geometric patterns in space and time.

\section{\textbf{References}}

\noindent Balter, M. 2010. Did working memory spark creative culture? \textit{Science} 328: 160--163.

\noindent Bickerton, D. 1995. \textit{Language and Human Behavior}. Seattle, WA: University of Washington Press.

\noindent Bouwmeester, D.; Pan J. W.; Mattle, K.; Eibl, M.; Weinfurter, H.; and Zeilinger, A. 1997. Experimental quantum teleportation. \textit{Nature} 390(6660): 575-579.    

\noindent Fuchs, C. A.; Mermin, N. D.; and Schack, R. 2013. An introduction to Qbism with an application to the locality of quantum mechanics. \textit{arXiv: 1311.5253v1 (quant-ph)}.

\noindent Han, C. J.; O’Tuathaigh, C. M.; van Trigt, L.; Quinn, J. J.; Fanselau, M. S.; Mongeau, R.; Koch, C.; and Anderson, D. J. 2003. Trace but not delay fear conditioning requires attention and the anterior cingulated cortex. \textit{Proc. Natl. Acad. Sci. USA} 100(22): 13087-13092.

\noindent Hackerm\"{u}ller, L.; Hornberger, K., Brezger, B.; Zeilinger, A.; and Arndt, M. 2004. Decoherence of matter waves by thermal emission of radiation. \textit{Nature 427}: 711-714. 

\noindent K\"{u}ppers, B.-O. 1990. \textit{Information and the Origin of Life}. Cambridge, Mass.: The MIT Press.

\noindent Kocsis, S.; Braverman, B.; Ravets, S.; Stevens, M. J.; Mirin, R. P.; Shalm, L. K.; and Steinberg, A. M. 2011. Observing the average trajectories of single photons in a two-slit interferometer. \textit{Science 332}: 1170-1173. 

\noindent Mermin, N. D. 2007. \textit{Quantum Computer Science}. Cambrisge University Press, New York.

\noindent Roederer, J. G. 1978. On the relationship between human brain functions and the foundations of physics. \textit{Found. of Phys. 8}: 423-438.

\noindent Roederer, J.G. 2003. Information and its role in nature. \textit{Entropy} 5: 1-31.

\noindent Roederer, J.G. 2004. When and where did information first appear in the Universe? In \textit{New Avenues in Bioinformatics} J. Seckbach et al., eds. Kluwer Acad. Publ., Dordrecht.

\noindent Roederer, J. G. 2005. \textit{Information and its Role in Nature}. Berlin, Heidelberg, New York: Springer-Verlag.

\noindent Roederer, J. G. 2012. Toward an information-based interpretation of quantum mechanics and the quantum-classical transition. {arXiv: 1108.0999v2 (quant-ph)}.

\noindent Schlosshauer, M. 2008. \textit{Decoherence and the Quantum-to-Classical Transition}. Berlin, Heidelberg: Springer-Verlag.

\noindent Shannon, C. E., Weaver W. W. 1949. \textit{The Mathematical Theory of Communication}. University of Illinois Press. 

\noindent Steane, A. M. 2007. Context, spacetime loops and the interpretation of quantum mechanics. \textit{J. Phys. A: Math. and Theor.} 40: 3223-3243. 

\noindent Tononi, G. and Koch, C. 2008. The neural correlates of consciousness: an update. \textit{Annals of the New York Academy of Science 1124}: 239-261.

\noindent Walker, S. I., Kim, H. and Davies, P. C. W. 2016. The informational architecture of the cell. In press \textit{Phil. Transactions}.  

\noindent Wheeler, J. A. 1989. Information, physics, quantum: The search for links. \textit{Proc. 3rd Int. Symp. Foundations of Quantum Mechanics}, Tokyo, 1989, pp.354-368.


\end{document}